\documentclass[10pt,letterpaper]{article}
\usepackage{opex3}
\usepackage{cite}
\usepackage{amsmath}
\begin{document}
\title{
Influence of large permanent dipoles on molecular orbital tomography
}
\author{Xiaosong Zhu,$^{1}$ Meiyan Qin,$^{1}$ Qingbin Zhang,$^{1,2}$ Yang Li,$^{1}$ Zhizhan Xu,$^{1,2}$ and Peixiang Lu$^{1,2}$$^{\star}$}
\address{$^1$ Wuhan National Laboratory for Optoelectronics and School of Physics, Huazhong University of
Science and Technology, Wuhan 430074, China\\
$^2$ Key Laboratory of Fundamental Physical Quantities Measurement of Ministry of Education, Wuhan 430074, China}
\email{$^\star$ lupeixiang@mail.hust.edu.cn}

\begin{abstract}
The influence of large permanent dipoles on molecular orbital tomography via high-order harmonic generation (HHG) is investigated in this work. It is found that, owing to the modification of the angle-dependent ionization rate resulting from the Stark shift, the one-side-recollision condition for the tomographic imaging can not be satisfied even with the few-cycle driving pulses. To overcome this problem, we employ a tailored driving pulse by adding a weak low-frequency pulse to the few-cycle laser pulse to control the HHG process and the recollision of the continuum electrons are effectively restricted to only one side of the core. Then we carried out the orbital reconstruction in both the length and velocity forms. The results show that, the orbital structure can only be successfully reproduced by using the dipole matrix elements projected perpendicular to the permanent dipole in both forms.
\end{abstract}

\section{Introduction}
The recent development in strong-field physics has opened wide perspectives for observing the structure and ultrafast electron dynamics in atoms and molecules with {\AA}ngstr\"{o}m and attosecond resolutions \cite{Lein,Lin,Smirnova,Pfeiffer,Zhou}. One of the most fascinating topics at the frontier of this area is the molecular orbital tomography (MOT) based on high-order harmonic generation (HHG), which was first proposed by J. Itatani \emph{et al} in \cite{Itatani}. By applying this method to a series of aligned molecules, they accomplished a tomographic reconstruction of the highest occupied molecular orbital (HOMO) of N$_2$. Since then, the MOT has attracted a great deal of attention for its great significance in uncovering the basic properties of molecules and chemical reactions \cite{Haessler,Salieres,Patchkovskii,Haessler2,Hijano,Vozzi,Diveki}. In most of the works, the target molecular orbitals are symmetric with either gerade or ungerade symmetry, while the MOT of nonsymmetric orbitals is more difficult. It has been discussed by Van der Zwan \emph{et al} \cite{Zwan} that the MOT of orbitals with arbitrary symmetry requires one to restrict the recollision of the continuum electrons to only one side of the core in the HHG process \cite{Corkum,Lewenstein}. Therefore, they suggested using extremely short few-cycle laser pulses to ensure the one-side-recollision and the molecular orbital of HeH$^{2+}$ was successfully reconstructed in their simulation. Besides, one can also consider using an $\omega+2\omega$ two-color laser pulse for the MOT of nonsymmetric orbitals \cite{Haessler,Qin}.

In addition to the arbitrary symmetry, another important characteristic of the nonsymmetric orbital is the permanent dipole. As a large permanent dipole would lead to an unnegligible Stark shift of the orbital energy, it would have influence on the MOT in two aspects. Firstly, it will cause additional laser-induce Stark phase acquired in the HHG of the nonsymmetric orbital. As the phases of field-free dipole matrix elements are needed in order to reconstruct molecular orbitals, one has to find a way to subtract the Stark phase from experimental measurements. The method is found in \cite{Etches}, in which the authors derived an analytical expression of the first-order Stark phase only depending on the measurable quantities. Secondly, previous works have shown that the Stark shift of the orbital energy can efficiently modify the angular dependence of the ionization rate and the distribution of the ionized electrons \cite{Holmegaard,Abu,Zhu}. This effect will make it more complicated to control the continuum electron for one-side-recollision and to reconstruct the orbital. This influence has however not been discussed before to the best of our knowledge and requires more concerns. Moreover, the influence of the nonsymmetric electron distribution on the reconstructed result in MOT has not been extensively discussed and calls for more studies too.

In this paper, the influence of large permanent dipoles on the MOT of nonsymmetric molecular orbital is investigated. It is found that, the one-side-recollision condition in HHG can not be satisfied even with the few-cycle laser pulses, owing to the modification of the angle-dependent ionization rate induced by the Stark shift. We employ a tailored laser pulse to overcome this problem and the orbital reconstruction is carried out in both the length and velocity forms. The reconstructed results are compared to the \emph{ab initio} orbital and the influence of the nonsymmetric electron distribution on the reconstructed result in MOT is discussed at the end.

\section{Theoretical model}
This section includes two independent parts: ``Simulating the HHG" and ``Reconstruction method". In the first part, we describe the modified Lewenstein model in the length form, which is employed in this work to simulate the HHG. In the second part, we introduce the orbital reconstruction method in both the length and velocity forms. Note that in the second part the inputting high-order harmonic spectra for the reconstruction process could be obtained from any ways: from calculations of different theories and forms or from experimental measurements.

\subsection{Simulating the HHG}
According to the Lewenstein model \cite{Lewenstein} and using the approximate saddle points \cite{Ivanov,Chirila}, the time-dependent dipole moment is calculated by (atomic units are used throughout)
\begin{equation}
\mathbf{D}(t)=\sum_{t_d}{a_{ion}(t_d)a_{prop}(t_d,t)a_{rec}[\mathbf{k}(t_d,t)]}+c.c.,
\end{equation}
where $a_{ion}$, $a_{prop}$ and $a_{rec}$ are the amplitudes of tunneling ionization, propagation after tunneling and recombination. $t_d$ is the approximate saddle time determined by solving
\begin{equation}
\frac{1}{2}[\mathbf{p}_{st}(t_d,t)+\mathbf{A}_f(t_d)]^2=0.
\end{equation}
$\mathbf{A}_f(t)$ is the vector potential corresponding to the electric field $\mathbf{F}(t)$, and $\mathbf{p}_{st}$ is the stationary momentum calculated by
\begin{equation}
\mathbf{p}_{st}(t_d,t)=-\frac{1}{t-t_d}\int_{t_d}^t{\mathbf{A}_f(t')dt'}.
\end{equation}
For atoms the amplitudes in Eq. (1) are given by
\begin{eqnarray}
\mathbf{a}_{ion}(t_d)&=&\sqrt{2I_p}\exp{[-\frac{(2I_p)^{\frac{3}{2}}}{3|F(t_d)|}]},\\
\mathbf{a}_{prop}(t_d,t)&=&(\frac{2\pi}{t-t_d})^{\frac{3}{2}} \frac{(2I_p)^{\frac{1}{4}}}{|F(t_d)|} \exp{\left[-i\int_{t_d}^t{dt'\frac{1}{2}\{[\mathbf{p}_{st}+\mathbf{A}_f(t')]^2+I_p}\}\right]},\\
\mathbf{a}_{rec}(\mathbf{k})&=&<\Psi_g|\mathbf{r}|e^{i\mathbf{k}\cdot\mathbf{r}}>.
\end{eqnarray}
$I_p$ is the ionization energy and $\Psi_g$ is the ground state in the form of hydrogenlike atom \cite{Lewenstein}. $\mathbf{k}=\mathbf{p}_{st}+\mathbf{A}_f(t)$ is the momentum of the return electron at the instant of recombination.

For molecules, the ionization amplitude $\mathbf{a}_{ion}$ is calculated with the MO-ADK theory \cite{Tong}, which takes into account the structure features of molecular orbitals. $\Psi_g$ is the HOMO of the molecule obtained from Gaussian 03 \emph{ab initio} code \cite{Gaussian}. For polar molecules whose HOMO has an unnegligible permanent dipole $\mathbf{\mu}_h$, the ionization energy becomes time-dependent in response to the external electric field \cite{Abu,Zhu,Etches2}:
\begin{equation}
I_p(t)=I_{p0}+\mathbf{\mu}_h\cdot\mathbf{F}(t),
\end{equation}
where $I_{p0}$ is the field free ionization energy. The permanent dipole $\mathbf{\mu}_h$ is calculated by
\begin{equation}
\mathbf{\mu}_h=-\int d\mathbf{r}\mathbf{r}\rho^H(\mathbf{r}).
\end{equation}
$\rho^H(\mathbf{r})$ is the electron density of the HOMO calculated by $\rho^H(\mathbf{r})=|\Psi_g(\mathbf{r})|^2$.

For the Lewenstein model in the length form, a modification should be applied in Eq. (6) that the momentum $\mathbf{k}$ is replaced by the effective momentum \cite{Kanai,Chen}
\begin{equation}
\mathbf{k}_{eff}=\sqrt{k^2+2I_p\gamma(k)}\frac{\mathbf{k}}{|k|}.
\end{equation}
This modification implies that the recombination occurs within the potential well and the acceleration effect of the binding potential is considered in this modified Lewenstein model \cite{Lein2,Kamta}.

In Eq. (9), the factor $\gamma(k)$ is given by
\begin{equation}
\gamma(k)=\left\{
\begin{array}{c}
1,\\
\\
\sin^2(\frac{\pi}{2}\frac{k^2}{I_p}).
\end{array}
\begin{array}{c}
k^2\geq I_p \\
\\
k^2<I_p
\end{array}
\right.
\end{equation}
This k-dependent factor avoids the unreasonable fact that all the return electrons have momenta larger than $\sqrt{2I_p}$. The introduction of $\gamma(k)$ also suppresses the strong background noises resulting from the jump of $\mathbf{k}_{eff}$ over the gap from $-\sqrt{2I_p}$ to $\sqrt{2I_p}$. As $\gamma=1$ for $k^2\geq I_p$, this term only affects the HHG of very low orders.

Finally, the harmonic spectrum is obtained by Fourier transforming the time-dependent dipole acceleration $\mathbf{a}(t)$:
\begin{eqnarray}
&\mathbf{E}_{XUV}=\int{\mathbf{a}(t)\exp(-iq\omega_{L}t)}dt,\\
&S_{q}=\mid\mathbf{E}_{XUV}\mid^{2},
\end{eqnarray}
where $\mathbf{a}(t)=\ddot{\mathbf{D}}(t)$, $\omega_{L}$ is the frequency of the driving pulse and $q$ corresponds to the harmonic order.

\subsection{Reconstruction method}
As shown in Fig. 1, (x,y,z) is the molecular frame and (x',y',z') is the laboratory frame. The driving laser field is linearly polarized along the x' axis and propagates along the z' axis. To reconstruct the molecular orbital, one must calculate or measure the high-order harmonic spectra (including the intensity and phase) from the oriented target molecule at various orientation angles $\theta$. In addition, one should also calculated or measure the high-order harmonic spectrum from the reference atom with the ionization energy similar to that of the target molecule. Then the recombination dipole moment $\mathbf{d}^L$ and the dipole velocity $\mathbf{d}^V$ for the molecular orbital in both x' and y' components are obtained by \cite{Haessler}
\begin{align}
d_{x'/y'}^{L/V}(\omega,\theta)&=\frac{1}{\eta(\theta_i)}\frac{E_{x'/y'}^{mol}(\omega,\theta)\exp[-i\Phi_{Stark}^{(1)}(\omega,\theta)]}{E^{ref}(\omega)}d_{ref}^{L/V}(k)\\
&=\frac{1}{\eta(\theta_i)}\frac{A_{x'/y'}^{mol}(\omega,\theta)}{A^{ref}(\omega)}\exp[i\varphi_{x'/y'}^{mol}(\omega,\theta)-i\Phi_{Stark}^{(1)}(\omega,\theta)-i\varphi^{ref}(\omega)] d_{ref}^{L/V}(k),
\end{align}
where E, A, $\varphi$ denote the complex electric field, amplitude and phase of the high-order harmonic radiation respectively. $\eta$ is the scaling factor in molecules depending on the ionization angle $\theta_i$ calculated by the MO-ADK theory. $\omega=q\omega_L$ is the harmonic frequency of the q-th order. The relation between $\omega$ and $k$ is determined by the energy conversation $\omega=k^2/2$. To subtract the additional acquired Stark phase from $E^{mol}$, the first-order Stark phase $\Phi_{Stark}^{(1)}(\omega,\theta)$ is introduced in Eqs. (13) and (14), which could be approximately evaluated with the analytical expression derived in \cite{Etches}. In the present paper, to focus our investigation on the influence of the permanent dipole in the aspect of modifying the ionization rate, the MOT is performed with the Stark phase perfectly subtracted in the simulation. In addition, the result of MOT with the Stark phase subtracted following \cite{Etches} as well as the influence of the permanent dipole in phase are briefly discussed at the end of Section 3.

The dipole matrix elements of the ``known'' reference atom can be obtained by
\begin{eqnarray}
d_{ref}^L(\mathbf{k})&=&-i\partial_\mathbf{k}\tilde{\Psi}_{ref}(\mathbf{k}),\\
d_{ref}^V(\mathbf{k})&=&\mathbf{k}\tilde{\Psi}_{ref}(\mathbf{k}),
\end{eqnarray}
where $\tilde{\Psi}_{ref}(\mathbf{k})$ is the Fourier transform of the ground state of the reference atom $\Psi_{ref}(\mathbf{r})$.

\begin{figure}[htb]
\centerline{
\includegraphics[width=8cm]{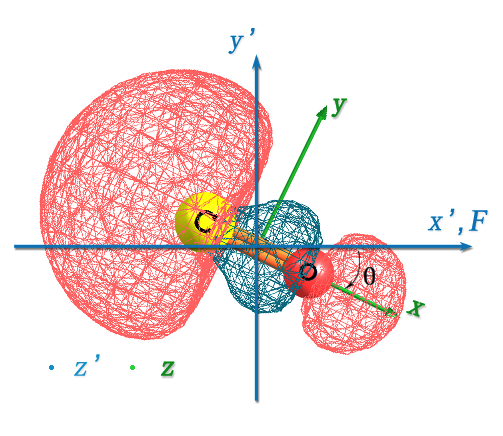}}
\caption{Definition of the reference frames in the MOT scheme: (x,y,z) is the molecular frame and (x',y',z') is the laboratory frame. The driving laser field $F$ is linearly polarized along the x' axis and propagates along the z' axis. $\theta$ is defined as the angle between x and x' axes. }
\end{figure}

By using the dipole moment or the dipole velocity, the reconstruction of the molecular orbital can be performed in the length form or in the velocity form.

In the length form, the orbital can be reconstructed as:
\begin{eqnarray}
&&\Psi_x^L(x,y)=\frac{1}{x}\sum_{\theta}\sum_q d_x^L(\omega,\theta)\exp\{i[k(\omega)\cos(\theta)x+k(\omega)\sin(\theta)y]\},\\
&&\Psi_y^L(x,y)=\frac{1}{y}\sum_{\theta}\sum_q d_y^L(\omega,\theta)\exp\{i[k(\omega)\cos(\theta)x+k(\omega)\sin(\theta)y]\},
\end{eqnarray}
where
\begin{eqnarray}
&&d_x^L(\omega,\theta)=d_{x'}^L(\omega,\theta)\cos(\theta)-d_{y'}^L(\omega,\theta)\sin(\theta),\\
&&d_y^L(\omega,\theta)=d_{x'}^L(\omega,\theta)\sin(\theta)+d_{y'}^L(\omega,\theta)\cos(\theta).
\end{eqnarray}
$\Psi_x^L(x,y)$ is the reconstructed orbital by using the dipole moments projected on the x axis of the molecular frame and $\Psi_y^L(x,y)$ is the reconstructed orbital by using the dipole moments projected on the y axis. For symmetric molecules, the same orbital can in principle be reconstructed from both the x and y components of the dipole moments ($d_x^L$ and $d_y^L$). However, due to the limited discrete sampling in Fourier space, they will most likely not give the same result \cite{Haessler}. Generally, the final reconstructed orbital is defined as
\begin{equation}
\Psi^L(x,y)=\frac{1}{2}[\Psi_x^L(x,y)+\Psi_y^L(x,y)]
\end{equation}
to hopefully average out the distortions.

Similarly, one can also reconstruct the orbital in the velocity form using the two projected components of the dipole velocity respectively:
\begin{eqnarray}
&&\Psi_x^V(x,y)=\sum_{\theta}\sum_q \frac{d_x^V(\omega,\theta)}{k(\omega)\cos(\theta)}\exp\{i[k(\omega)\cos(\theta)x+k(\omega)\sin(\theta)y]\},\\
&&\Psi_y^V(x,y)=\sum_{\theta}\sum_q \frac{d_y^V(\omega,\theta)}{k(\omega)\sin(\theta)}\exp\{i[k(\omega)\cos(\theta)x+k(\omega)\sin(\theta)y]\},
\end{eqnarray}
with
\begin{eqnarray}
&&d_x^V(\omega,\theta)=d_{x'}^V(\omega,\theta)\cos(\theta)-d_{y'}^V(\omega,\theta)\sin(\theta),\\
&&d_y^V(\omega,\theta)=d_{x'}^V(\omega,\theta)\sin(\theta)+d_{y'}^V(\omega,\theta)\cos(\theta).
\end{eqnarray}

\section{Result and discussion}
This section consists of two parts. In the first part, we analyze the HHG of a polar molecule to demonstrate the influence of the large permanent dipole on MOT and suggest a method to overcome this unfavorable influence. In the second part, we compare and discuss the reconstructed results in different forms and by using different projected components of the dipole matrix elements.

\subsection{Simulating and controlling the HHG}
In this work, we apply CO as the target molecule. The HOMO of CO has a large permanent dipole of 1.72 a.u. pointing from C to O nucleus. The ionization energy of CO is 0.5152 a.u., and the reference atom is Kr with the ionization energy of 0.5145 a.u. We apply a linearly polarized three-cycle sin$^2$ driving pulse $F(t)=F_0\sin^2(\frac{\pi t}{3T_0})\cos(\omega_Lt+\varphi_0)$, where $T_0$ is the optical cycle and $\varphi_0=1.25\pi$. This form of laser pulse (with an opposite sign) has been used to accomplish the MOT of the nonsymmetric molecule HeH$^{2+}$ \cite{Zwan}. The intensity and wavelength of the laser pulse we use are $1\times10^{14}$ W/cm$^2$ and 1.5 $\mu$m respectively. Employing the 1.5 $\mu$m mid-IR laser source allows us to extend the harmonic spectrum considerably with a low laser intensity below the barrier suppression intensity \cite{Vozzi,Vozzi2}, which ensures the validity of the used MO-ADK model and strong-field approximation (SFA) in our theoretical description. Another benefit of keeping the intensity low is that it will reduce the effect of the Stark shift.
\begin{figure}[htb]
\centerline{
\includegraphics[width=10.5cm]{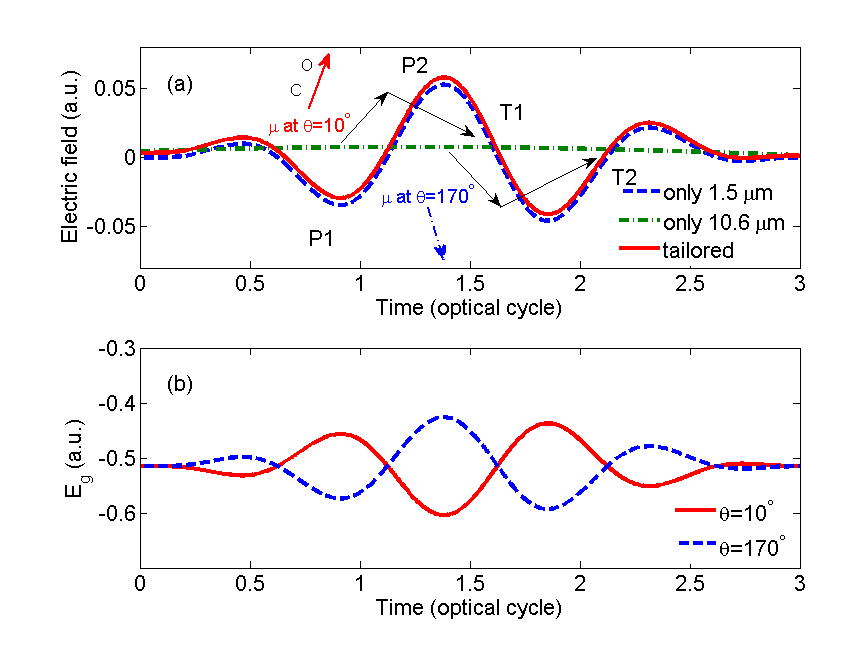}}
\caption{(a) The electric fields of the driving, assistant and tailored pulses. The black arrows illustrate the two main contributing sets of trajectories in the HHG: the electron of P1$\rightarrow$T1 trajectory returns to the core from +x' position with a negative momentum, while the electron of P2$\rightarrow$T2 trajectory returns from the -x' side with positive momentum. (b) The time-dependent orbital energy $E_g=-I_p$ for $\theta=10^\circ$ and $170^\circ$ respectively.}
\end{figure}

The electric field of the driving pulse is depicted by the dashed blue curve in Fig. 2(a). The HHG process is also illustrated by the black arrows in Fig. 2(a): under this three-cycle laser pulse, the HHG is dominantly contributed by the electrons ionized by the two peaks P1 and P2. For the electrons ionized by P1, they first drift towards the positive direction and finally return to the core at times around T1 with negative momenta, emitting the high-order harmonic photons. On the contrary, the electrons ionized by P2 finally return to the core at times around T2 with positive momenta.

When the effect of the permanent dipole of CO is ignored, the ionization energy is constant. The ionization probabilities in the +x' and -x' directions are dominantly determined by the waveform of the electric field, due to the exponential dependence of the tunneling ionization rate on the amplitude of electric field. Therefore, as the maximum amplitude of P2 is bigger than that of P1, electrons are more easily to be ionized by the P2 peak and the high-order harmonic radiation will be stronger at T2 than at T1. This is confirmed by our calculation shown in the first row in Fig. (3). In Fig. 3(a), the red and green solid curves show the x' and y' components of the high-order harmonic spectrum from CO oriented at $\theta=10^\circ$ without taking into account the Stark shift. The harmonic spectrum from the reference atom is also shown in Fig. 3(a) by the dashed curve. Figures 3(b) and (c) show the corresponding time-frequency analyses \cite{Antoine,Etches2} of the molecular HHG in the parallel (x') and perpendicular (y') directions respectively. It is found that the atomic spectrum exhibits double-plateau structure, while only the plateaus on the lower energy side are observed in the molecular spectra. This is because the plateaus on the higher energy side are so weak as to be below the background noise. The result implies that the plateaus on the lower energy side are at least 3-4 orders of magnitude higher than those on the higher energy side in both x' and y' components in the spectrum from CO. According to the time-frequency analyses, the plateaus on the lower energy side are mainly contributed by the Q2 peaks, which are generated by the recollision of electrons with positive momenta at T2. While the plateaus on the higher energy side are contributed by the Q1 peaks, which are generated by the recollision of electrons with negative momenta at T1. From both the observations in the spectra (Fig. 3(a)) that the plateaus on the lower energy side are much higher than those on the higher energy side and in the time-frequency distributions (Figs. 3(b) and (c)) that Q2 are much stronger than Q1, it can be concluded that the high-order harmonics are generated by the recollision of electrons from only one side.

\begin{figure}[htb]
\centerline{
\includegraphics[width=13cm]{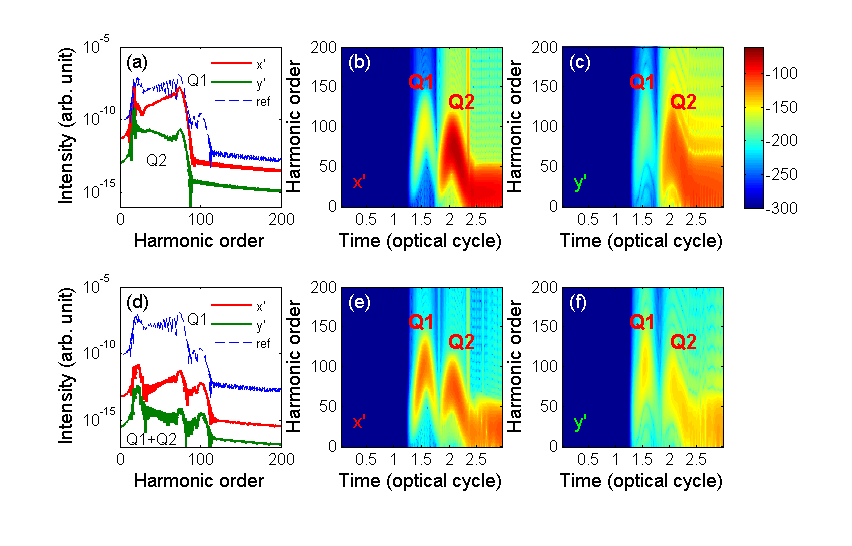}}
\caption{ First row: (a) the x' and y' components of the high-order harmonic spectrum from the target molecule CO and (b)(c) the time-frequency distributions for each component. The HHG is calculated in response to the few-cycle pulse neglecting the effect of the permanent dipole. Second row: (d) the x' and y' components of the molecular harmonic spectrum and (e)(f) the time-frequency distributions for each component, after taking into account the Stark shift. The spectrum of the reference atom is also depicted by the dashed blue curve in panels (a) and (d). The values in the color bar correspond to $10log_{10}$(Intensity).}
\end{figure}

However, when the stark shift is considered, the one-side-recollision condition is worse satisfied. As shown in Figs. 3(d), the plateaus on the higher energy side appear and are in the same level with the plateaus on the lower energy side in the harmonic spectra of CO in both components. The time-frequency analyses also show that the peaks Q1 and Q2 have comparable intensities. It has been discussed that Q1 is generated by the electrons with negative return momenta and Q2 is generated by the electrons with positive momenta. Therefore, the results indicate that the electrons with both negative and positive momenta have comparable contributions to the HHG.

To explain why the Stark shift makes the one-side-recollision condition worse satisfied, the time-dependent orbital energy $E_g=-I_p$ in response to the external electric field $F(t)$ is plotted in Fig 2(b). The electron is more likely to be ionized with a high $E_g$ and is harder to be ionized when $E_g$ is low. Therefore, the ionization rate of CO is decreased when $\mu_h$ is parallel with the electric filed $F$ and is increased in the antiparallel geometry. In the case of 10$^\circ$ orientation, as shown by the red solid curve in Fig. 2(b), the ionization by P1 is enhanced by the Stark shift and the ionization by P2 is suppressed, leading to the comparable ionization probabilities from both peaks. And these comparable ionization probabilities finally lead to the comparable HHG contributions by electrons returning from opposite sides.

\begin{figure}[htb]
\centerline{
\includegraphics[width=10cm]{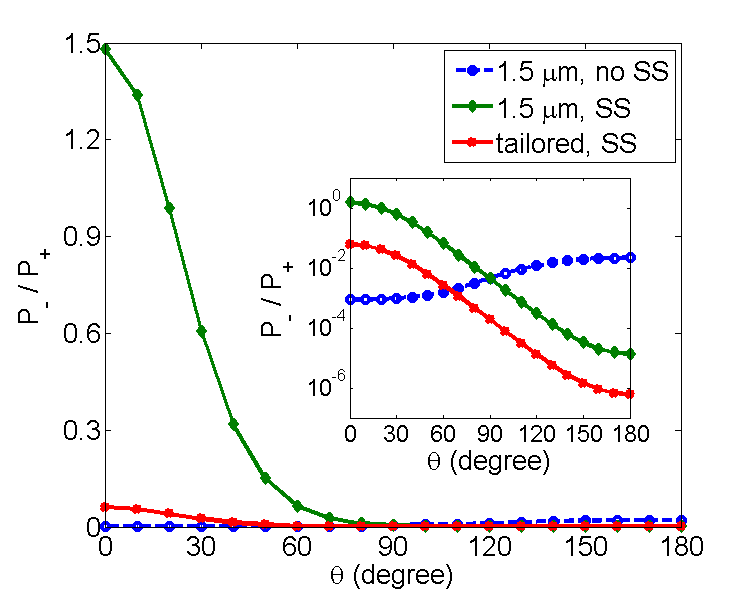}}
\caption{The ratio between the probabilities that an electron returns to the core with negative and positive momentum for various orientation angles in the cases of: (1) using only the three-cycle driving pulse without taking into account the Stark shift (SS) (blue circles), (2) using only the three-cycle driving pulse with the SS (green diamonds), and (3) using the tailored pulse with the SS (red pentagrams). The inset shows the ratios on a logarithmic scale.}
\end{figure}

As the MOT requires extracting structure information (the slices) of the target orbital from different directions, the one-side-recollision condition should be satisfied at various orientation angles. To evaluate the ratio between the probabilities that an electron returns to the core with negative and positive momenta at different angles, a semi-classical method introduced in \cite{Zwan2} is used. First of all the electrons are tunneling ionized with the probability calculated by the MO-ADK theory. After ionization, the freed electron follows a classical trajectory, and we collect the electrons which finally return to the core. Every classical return receives a weight based on the ionization probability and a factor of $\tau^3$, where $\tau$ is the time the electron spends in the continuum from ionization to recombination and reflects the effect of wave-packet spreading. After the collection, the return probability of the electron with negative momentum $P_-$ and the return probability of electron with positive momentum $P_+$ are obtained. The ratio of the probabilities $P_-/P_+$ for orientation angles from 0$^\circ$ to 180$^\circ$ is plotted in Fig. 4, where the blue circles and the green diamonds present the calculated results without and with the Stark effect taken into account respectively. Comparing the green diamonds to the blue circles, the Stark shift leads to more recollision with negative momentum especially for the orientation at small angles. The ratio of $P_-/P_+$ is bigger than 0.1 for $\theta<60^\circ$ and goes up to bigger than 1 for $\theta=0$ and $10^\circ$.

To overcome this problem, one may first think of using an even lower intensity for the driving pulse to further reduce the effect of the Stark shift. However, too low intensity will bring the problem of a bad signal to noise ratio. Instead, in previous experimental and theoretical works higher intensities than 1$\times10^{14}$ W/cm$^2$ were preferred, \emph{e.g.} Ref. \cite{Vozzi,Etches2}. Therefore, in this work we suggest a method to solve this problem by using a tailored laser pulse with the relatively high intensity. The tailored laser pulse is synthesized by adding a weak low-frequency laser field to the few-cycle laser field. The low-frequency field can be provided by different kinds of light sources such as the CO$_2$ laser \cite{Serrat,Luo} and the THz radiation \cite{Hong,Kovacs}. The basic idea of adding this assistant field is to enlarge the difference between the amplitudes of P1 and P2. In the present work, we apply the low-frequency pulse from CO$_2$ laser with the wavelength of 10.6 $\mu$m. The intensity of this pulse is 1\% of that of the few-cycle driving pulse and the electric field is $F'=F_0'\cos(\omega_L't+\varphi_0')$ with $\varphi_0'=1.7\pi$. The low-frequency field and the tailored field are presented by the dash-dotted green and solid red curves in Fig. 2(a) respectively.

\begin{figure}[htb]
\centerline{
\includegraphics[width=13cm]{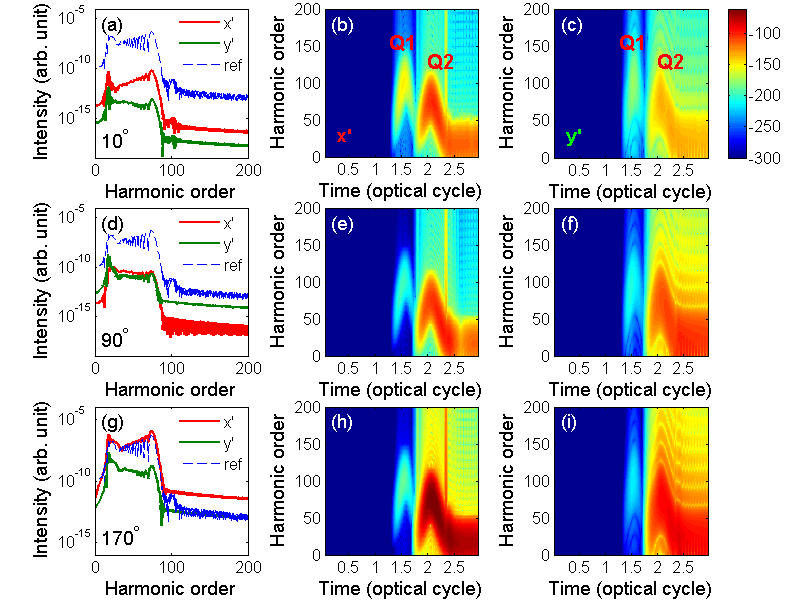}}
\caption{The calculated high-order harmonic spectra (the first column) and the corresponding time-frequency distributions for the x' and y' components of the molecular harmonic spectra (the second and third columns), by using the tailored pulse taking into account the Stark shift. The first, second and third rows correspond to $\theta=10^\circ$, $90^\circ$ and $170^\circ$ respectively. The values in the color bar correspond to $10log_{10}$(Intensity).}
\end{figure}

The ratio $P_-/P_+$ under this tailored field is presented by the red pentagrams in Fig. 4. Comparing to the green diamonds, the recollision with negative momentum is suppressed: the ratio is two orders of magnitude decreased and is below 0.1 for all the orientation angles (for $180^\circ<\theta<360^\circ$ the ratio at $\theta$ equals that at $2\pi-\theta$). The result shows that the tailored laser pulse can efficiently restrict the recollision of the continuum electrons to only one side of the core.

The high-order harmonic spectra at three typical orientation angles 10$^\circ$, 90$^\circ$ and 170$^\circ$ calculated with the quantum model described in Section 2.1 are shown in the first column in Fig. 5. The second and third columns present the time-frequency analyses of the molecular harmonic spectra in the x' and y' components respectively. These results are consistent with those from the semi-classical method. The peaks Q2 are much stronger than Q1 in all the time-frequency distributions. Regarding the molecular harmonic spectra, the plateaus on the lower energy side (predominantly contributed by electrons with positive momenta) are 3-4 orders higher than those on the higher energy side (contributed by electrons with negative momenta) at $\theta=10^\circ$ as shown in Fig. 5(a). And the plateaus on the lower energy side are at least 4 orders of magnitude higher than those on the higher energy side for $\theta=90$ and $170^\circ$ as shown in Figs. 5(d) and (g). Therefore, the one-side-recollision condition is met and the HHG by this tailored pulse is satisfying for the MOT of the nonsymmetric target orbital.

\begin{figure}[htb]
\centerline{
\includegraphics[width=12cm]{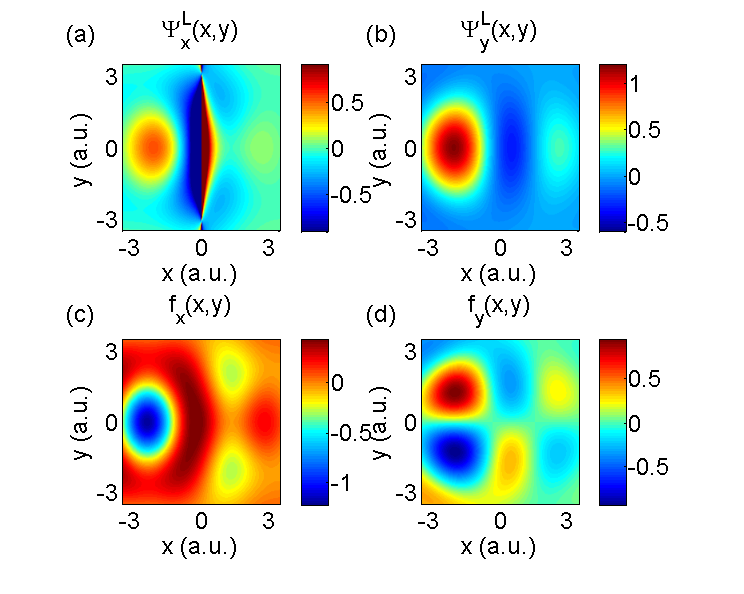}}
\caption{(a)(b) The orbitals reconstructed in the length forms by using the dipole moments projected on the x and y axis respectively. (c)(d) The intermediate functions $f_x(x,y)$ and $f_y(x,y)$.}
\end{figure}

\subsection{Tomographic reconstruction}
The MOT result in the length form is shown in Fig. (6). To reconstruct the orbital, we use the odd harmonics from the 23rd to the 79th orders (corresponds to 19-65 eV) generated from CO at $\theta$ from 0$^\circ$ to 360$^\circ$ with the angular step $\Delta\theta=10^\circ$. Figures 6(a) and (b) display the reconstructed orbitals $\Psi_x^L$ and $\Psi_y^L$ using the dipole moments projected on the x and y axes in the molecular frame respectively. The $\emph{ab initio}$ orbital is also shown in Fig. 7(c). Compared with Fig. 7(c), Fig. 6(b) successfully reproduces the main structure of the nonsymmetric target orbital, while Fig. 6(a) fails to reproduce the structure. There is an artificial structure close to x=0 in Fig. 6(a) with a sharp jump from the negative value to the positive value. This is due to the nonsymmetric distribution of the orbital.

To illustrate this, in Figs. 6(c) and (d) we present the intermediate functions $f_x$ and $f_y$, which are defined as \cite{Itatani}
\begin{equation}
f_\xi(x,y)=\sum_{\theta}\sum_q d_\xi^L(\omega,\theta)\exp\{i[k(\omega)\cos(\theta)x+k(\omega)\sin(\theta)y]\},\ \xi=x,y.
\end{equation}
The relation between $\Psi^L$ and $f$ is
\begin{equation}
\Psi_\xi^L=\frac{1}{\xi}f_\xi(x,y),\ \xi=x,y.
\end{equation}
Both $f_x(x,y)$ and $f_y(x,y)$ are symmetrically distributed along the y direction and are nonsymmetrically distributed along the x axis, corresponding to the nonsymmetric electric density distribution of the target orbital along the permanent dipole. Therefore, one will have no problem in dividing $f_y$ by $y$ to obtain $\Psi_y^L$, because the nodal plane of $f_y$ locates right at $y=0$. But the nodal plane is bent and not coincident with $x=0$ for $f_x$. As a result, numerical problem arises when dividing $f_x$ by $x$ with x$\rightarrow$0, which leads to the artificial structure and sharp jump in Fig. 6(a).

\begin{figure}[htb]
\centerline{
\includegraphics[width=12cm]{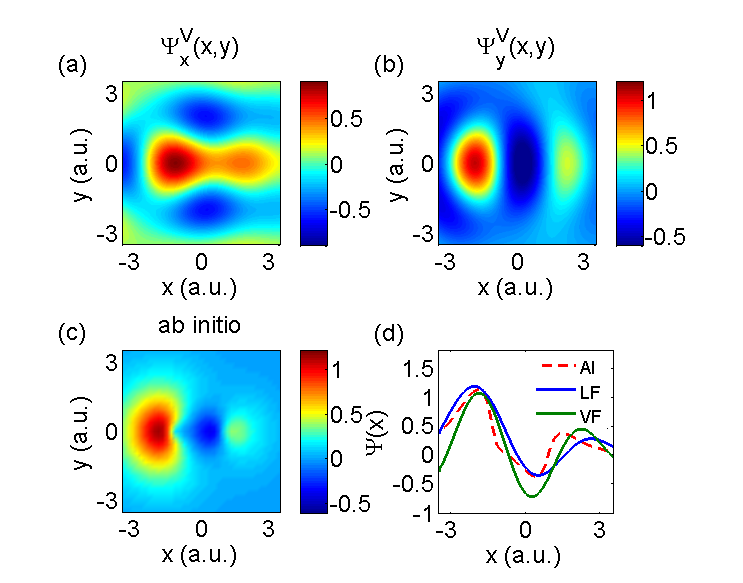}}
\caption{(a)(b) The orbitals reconstructed in the velocity form by using the dipole velocities projected on the x and y axis respectively. (c) The \emph{ab initio} orbital. (d) Slices of $\Psi_y^L$ (LF), $\Psi_y^V$ (VF) and the \emph{ab initio} orbital (AI) along the x axis.}
\end{figure}

The reconstruction is also performed in the velocity form as shown in Figs. 7(a) and (b). The scan of $\theta$ starts from 5$^\circ$ and increases in the step of $10^\circ$ to avoid $\cos(\theta)=0$ or $\sin(\theta)=0$ in the denominators in Eqs. (22) and (23). $\Psi_x^V$ and $\Psi_y^V$ are the reconstructed orbitals using the dipole velocities projected on the x and y axes respectively. Again, $\Psi_x^V$ and $\Psi_y^V$ give different results. $\Psi_y^V$ successfully reproduces the main structure of the target orbital, while $\Psi_x^V$ is incorrect. To further examine the influence of the nonsymmetric distribution of the orbital on the reconstructed result, we have also tried the MOT of another kind of orbital which is nonsymmetrically distributed along the y axis. On the contrary to the results shown in Figs. 6 and 7, $\Psi_x^L$ and $\Psi_x^V$ successfully reproduce the orbital this time, while $\Psi_y^L$ and $\Psi_y^V$ are incorrect (not shown). All the results indicate that, the MOT can only be accomplished by using the dipole matrix elements projected perpendicular to the permanent dipole in both the length and velocity forms for the nonsymmetric orbitals.

Further comparing the reconstructed orbitals $\Psi_y^L$ and $\Psi_y^V$ with the \emph{ab initio} orbital, it is found that there is still detailed structure missed in the reconstructed results near the left core. This is due to the limited spectral range ``detected'' for the MOT \cite{Qin}. In Fig. 7(d), we compare the slices along the x axis of $\Psi_y^L$, $\Psi_y^V$ and the \emph{ab initio} orbital. The comparison shows that, the result obtained in the length form agrees with the \emph{ab initio} orbital well for the positions and amplitudes of the left and middle peaks, but the position of the right peak deviates far away from the \emph{ab initio} orbital. The result obtained in the velocity form matches the \emph{ab initio} orbital better for the positions of all the three peaks, but the values at x=-3 and 0 a.u. deviate a lot from those of the \emph{ab initio} orbital. Both forms have limitation in reproducing the exact shape of the target orbital. Furthermore, the reconstruction of the orbital in the y direction is not satisfying. This error is also observed in the previous works especially for the $\sigma$ orbital. Besides the limited spectrum range, the imaging errors may also result from several other reasons, such as the discrete sampling in the frequency domain with the interval of $2\omega_L$, and the SFA and single-active-electron approximation used in the theoretical model \cite{Haessler,Salieres,Patchkovskii,Haessler2,Vozzi}. In the MOT process, when the Stark phase is approximately subtracted by using the analytical expression derived in \cite{Etches}, the main structure of the target orbital is also successfully reproduced in both the length and velocity forms. The reconstructed orbitals have similar shapes to those shown in Figs. 6 and 7, except that the whole structures are shifted towards the right in both forms. These shifts may therefore originate from the approximation in evaluating the Stark phase. On the other hand, this result implies that the influence of the permanent dipole in phase would lead to the shift of the reconstructed orbital, if the Stark phase is not exactly subtracted.

\section{Conclusion}
The influence of large permanent dipoles on the MOT is investigated in this paper. It is found that, owing to the modification of the angle-dependent ionization rate induced by the Stark shift of the orbital energy, the one-side-recollision condition can not be satisfied even with the few-cycle driving pulse. To overcome this problem, we employ a tailored laser pulse by adding a weak low-frequency pulse to the few-cycle pulse to control the HHG process and the requirement for the one-side-recollision is met. Then the reconstruction of the target orbital is performed in both the length and velocity forms. The results show that, the orbital structure can only be successfully reproduced by using the dipole matrix elements projected perpendicular to the permanent dipole in both forms.

\section*{Acknowledgment}
This work was supported by the National Natural Science Foundation of China under Grants No. 11234004 and 60925021, the 973 Program of China under Grant No. 2011CB808103 and the Doctoral fund of Ministry of Education of China under Grant No. 20100142110047.
\end{document}